\documentclass{article}
\usepackage{amsmath,amsthm}
\usepackage{amsmath,amssymb}

\numberwithin{equation}{section}

\def\Gg {{\cal G}} 
\def\Hg {{\cal H}} 
\def\Ig {{\cal I}} 
\def\Kg {{\cal K}} 
\def\Sg {{\cal S}} 
\def\Tg {{\cal T}} 
\def\Vg {{\cal V}} 
\def\Xg {{\cal X}} 
\def\Yg {{\cal Y}} 
\def\R{{\mathbb R}}

\def\C{{\mathbb C}}

\def\ux{{\underline{x}}} 
\def\uy{{\underline{y}}}

\def\Bb{{\bf B}}

\def\hs{\hslash}

\def\a{\alpha}
\def\b{\beta}
\def\c{\gamma}
\def\Ga{\Gamma}
\def\d{\delta}

\def\ep{\varepsilon}

\def\th{\theta}

\def\l{\lambda}
\def\L{\Lambda}
\def\m{\mu}

\def\r{\rho}
\def\s{\sigma}
\def\Si{\Sigma}
\def\t{\tau}
\def\w{\omega}

\def\lbeq(#1){\label{eqn:#1}}
\def\refeq(#1){{\rm (\ref{eqn:#1})}}
\def\lbth(#1){\label{th:#1}}
\def\refth(#1){{\rm Theorem \ref{th:#1}}}
\def\refths(#1,#2){{\rm Theorems \ref{th:#1} and \ref{th:#2}}}
\def\refthb(#1){{\bf Theorem \ref{th:#1}}}
\def\lblm(#1){\label{lm:#1}}
\def\reflm(#1){{\rm Lemma \ref{lm:#1}}}
\def\lbprp(#1){\label{prp:#1}}
\def\refprp(#1){{\rm Proposition \ref{prp:#1}}}
\def\reflmb(#1){{\bf Lemma \ref{lm:#1}}}
\def\lbcor(#1){\label{cor:#1}}
\def\refcor(#1){{\rm Corollary \ref{cor:#1}}}
\def\lbrm(#1){\label{rm:#1}}
\def\refrm(#1){{\rm Remark \ref{rm:#1}}}
\def\lbass(#1){\label{ass:#1}}
\def\refass(#1){{\rm Assumption \ref{ass:#1}}}
\def\refasss(#1, #2){{\rm Assumptions \ref{ass:#1} 
and \ref{ass:#2}}}
\def\refassss(#1, #2, #3){{\rm Assumptions \ref{ass:#1},\ 
\ref{ass:#2} and \ref{ass:#3}}}
\def\lbdf(#1){\label{df:#1}}
\def\refdf(#1){{\rm Definition \ref{df:#1}}}
\def\lbsec(#1){\label{s:#1}}
\def\refsec(#1){{\rm \S\ref{s:#1}}}
\def\lbsubsec(#1){\label{ss:#1}}
\def\refsubsec(#1){{\rm \S\ref{ss:#1}}}

\def\bgth{\begin{theorem}}
\def\edth{\end{theorem}}
\def\bgprp{\begin{proposition}}
\def\edprp{\end{proposition}}
\def\bgdf{\begin{definition}}
\def\eddf{\end{definition}}
\def\bglm{\begin{lemma}}
\def\edlm{\end{lemma}}
\def\bgcor{\begin{corollary}}
\def\edcor{\end{corollary}}
\def\bgpf{\begin{proof}}
\def\edpf{\end{proof}}
\def\bgrm{\begin{remark}}
\def\edrm{\end{remark}}
\def\bgass{\begin{assumption}}
\def\edass{\end{assumption}}

\def\bqn{\begin{equation}}
\def\eqn{\end{equation}}

\def\ben{\begin{enumerate}}
\def\een{\end{enumerate}}
\def\ep{\varepsilon}
\def\ph{\varphi}

\def\la{\langle}
\def\ra{\rangle}

\def\ax{{\la x \ra}}

\def\br{\begin{array}}
\def\er{\end{array}}
\def\lap{\Delta}
\newcommand {\pa}{\partial}

\newtheorem{theorem}{Theorem}[section]
\newtheorem{lemma}[theorem]{Lemma}
\newtheorem{proposition}[theorem]{Proposition}
\newtheorem{definition}[theorem]{Definition}
\newtheorem{corollary}[theorem]{Corollary}

\newtheorem{assumption}[theorem]{Assumption}
\theoremstyle{definition}
\newtheorem{remark}[theorem]{Remark}

\def\loc{{\rm loc}}

\title{Existence and regularity of propagators for 
multi-particle Schr\"odinger equations in external 
fields}
\author{
K. Yajima\thanks{
Department of Mathematics, Gakushuin University, 
1-5-1 Mejiro, Toshima-ku, Tokyo 171-8588, Japan. 
}
}
\date{}
\begin{document}
\allowdisplaybreaks
\maketitle

\begin{abstract}
We consider Schr\"odinger equations for $N$ number of particles 
in (classical) electro-magnetic fields which are interacting 
each other via time dependent inter-particle potentials. 
We prove that they uniquely generate unitary propagators 
$\{U(t,s), t,s \in \R\}$ on the state space $\Hg$ under the conditions 
that fields are spatially smooth and do not grow too rapidly at infinity 
so that propagators for single particles satisfy Strichartz  
estimates locally in time and, that local singularities of 
inter-particle potentials are not too strong that time 
frozen Hamiltonians define natural selfadjoint realizations 
in $\Hg$. We also show, under very mild 
additional assumptions on the time derivative of  
inter-particle potentials, that propagators possess 
the domain of definition of the quantum harmonic oscillator 
$\Si(2)$ as an invariant subspace such that, for initial 
states in $\Si(2)$, solutions are $C^1$ functions of 
the time variable with values in $\Hg$. 
New estimates of Strichartz type for propagators  
for $N$ independent particles in the field will be proved 
and used in the proof. 
\end{abstract}

\section{Introduction} 

We consider $N$ number of $d$-dimensional non-relativistic 
quantum particles of masses $m_j>0$ and charges $e_j\in \R$, 
$0\leq j \leq N$. We denote  the position of $j$-th 
particle by $x_j=(x_{j1},\dots,x_{jd})\in \R^d$, $dx_j$ is 
the $d$-dimensional Lebesgue measure 
and $\ux=(x_1, \dots, x_N)\in \R^{Nd}$. 
If we ignore the spin and statistics, 
the state of the particles is described by the 
unit ray of the Hilbert space 
\[
\Hg= L^2(\R^{Nd})=\left\{u(\ux) \colon 
\|u\|^2=
\int_{\R^{Nd}}|u(x_1, \dots, x_N)|^2 dx_1 \dots dx_N<\infty 
\right\}.   
\]
We consider the situation that the particles are placed in 
the (classical) electro-magnetic field described by the 
electric scalar and the magnetic vector potentials given 
respectively by $\ph(t,x)$ and 
$A(t,x)=(A_1(t,x), \dots, A_d(t,x))$, 
$(t,x) \in \R \times \R^d$ and they are interacting each 
other via the inter-particle forces given by the 
potential $V(t,\ux)$. 
If single-particle forces from additional external 
sources are present, we include them into $V(t,\ux)$. 
Then, the Hamiltonian of the system is given by 
\bqn \lbeq(2)
H(t)= \sum_{j=1}^N 
\left(\frac1{2m_j}(-i\hs\nabla_j - e_j A(t,x_j))^2 + 
e_j \ph(t,x_j)\right) + V(t,\ux), 
\eqn 
where 
$\nabla_j= (\pa/\pa x_{j1},\dots,\pa/\pa x_{jd})$ 
for $1\leq j \leq N$, $\hs=h/2\pi$ and $h$ is the Planck 
constant and, the dynamics is 
governed by the Schr\"odinger equation   
\bqn 
i\hs\frac{du}{dt} = H(t)u(t) \lbeq(1) 
\eqn 
for $\Hg$-valued function 
$u(t)=u(t,\ux)$ of $t \in \R$. Hereafter 
we set $\hs=1$. 

In this paper, we prove under rather general assumptions 
on the potentials which will be made precise below that 
Eqn. \refeq(1) generates a unique dynamics of the 
particles, or it uniquely generates a strongly 
continuous family of unitary operators 
$\{U(t,s) \colon -\infty<t,s<\infty\}$ on $\Hg$ such that 
$u(t)=U(t,s)f$ for $f\in \Hg$ produces the solution 
of \refeq(1) which satisfies the initial condition 
$u(s)=f$. We call $\{U(t,s)\}$ the {\it unitary propagator} 
for \refeq(1). It satisfies the Chapman-Kolmogorov equation 
\bqn \lbeq(CK)b
U(t,s)= U(t,r) U(r,s), \quad U(t,t)={\bf 1}, 
\quad t,s,r \in \R, 
\eqn 
where ${\bf 1}$ is the identity operator on $\Hg$; 
we also prove, under slightly stronger assumptions, 
that $\{U(t,s)\}$ possesses the domain of 
definition $\Si(2)$ of the quantum harmonic 
oscillator as an invariant subspace such that 
$u(t)=U(t,s)f$ with $f \in \Si(2)$ is 
$C^1$ function of $t$ with values in $\Hg$ and it 
satisfies \refeq(1) as an evolution equation in $\Hg$. 

This is an improvement and an extension to $N$-particle 
systems of author's previous papers \cite{Y-1,Y-2} on 
the same subject for the case $N=1$ and, before stating 
the main theorems, we think it 
appropriate to briefly touch upon the history of the subject. 

The existence and the uniqueness of the unitary propagator 
for Schr\"odinger equations is certainly one of the most 
fundamental and the oldest problems in mathematics for 
quantum mechanics and it has been intensively and deeply 
studied by many authors since its advent (cf. \cite{Dirac}). 
If the Hamiltonian $H(t)=H$ is 
independent of time, the problem is virtually equivalent 
to the selfadjointness of $H$ and, after a long and 
extensive study by various authors since Kato's seminal 
paper \cite{Ka-F}, it is now considered that the problem 
has almost been settled (see e.g. \cite{CFKS}, \cite{RS2} 
and reference therein for a large and rich literature). 

For time dependent Hamiltonians $H(t)$, many and various 
methods have likewise been invented by many authors 
for producing the unitary propagator for \refeq(1). 
Adaptations of energy methods for the Cauchy problem for 
hyperbolic equations (e.g. \cite{LM, Taylor, Ichi}), 
the method of parabolic regularization (\cite{LM}) and 
the application of the theory of temporally inhomogeneous 
semi-groups (e.g. \cite{Yo,RS2,K-1,K-2}), which we simply 
call semigroups in the sequel, to mention a few. 

Among these, we think that the application 
of semigroup theory is the simplest and the furthest reaching, 
particularly for multi-particle Schr\"odinger equations   
and, most authors refer to this method for the existence 
of propagators. This theory, however, requires conditions 
like {\it $D(H(t))$ is $t$-independent and 
$\pa_t H(t)$ is $H(t)$- or $H(t)$-form bounded} 
or similar ones when $D(H(t))$ is $t$-dependent, 
which often impose rather strong restrictions on 
potentials, see e.g. \cite{Y-3, AY} where the existence 
of a unique unitary propagator for the case $N=1$ is 
proved when potentials satisfy conditions which are almost 
necessary for $H(t)$ to be selfadjoint but under rather 
strong assumptions on the time derivative. 

To see that the lack of this property can lead to the breakdown 
of the uniqueness of the propagator, 
we consider the following example: 
\bqn \lbeq(non-1)
i \pa_t u = \frac12(-i\nabla + {\s}tx\ax^{\s-2})^2 u 
+  C \ax^{\s}u 
= H_C(t)u, \quad (t,x) \in \R \times \R^d,  
\eqn 
where $\s\geq 0$ and $C>0$. The operator $H_C(t)$ is 
seladjoint on $\Hg$ with maximal domain and is unitarily equivalent 
via $T(t)u(x) =  e^{it\ax^\s}u(x)$ to 
\bqn \lbeq(unit-eq)
T(t)^\ast H_C(t) T(t)=  -\frac12\lap + C \ax^{\s} 
\equiv H_{C,0}. 
\eqn 
Thus, $\pa_t H_{C}(t)= -\frac{i {\s}}{2} 
T(t)\left(x\ax^{\s-2}\cdot \nabla + 
\nabla\cdot x\ax^{\s-2}\right) T(t)^\ast$  
is {\it not} bounded with respect to $H_C(t)$ if $\s>2$ for any $C>0$ 
(but it is if $\s\leq 2$). On the other hand, the change of gauge  
$v(t,x) = T(t) u(t,x)$ transforms \refeq(non-1) into  
\bqn \lbeq(non-a)
i\pa_t v(t) = \left(-\frac12 \lap +(C-1)\ax^{\s}\right)v(t)
= H_{C-1, 0}v(t)  
\eqn 
and, as is well known (\cite{RS2}), $H_{C-1, 0}$ is not essentially 
selfadjoint on $C_0^\infty(\R^d)$, if $\s>2$ and $C<1$, and has an 
infinitely number of selfadjoint extensions $\{H_\l\colon \l \in \L\}$ 
each of which generates different dynamics 
for the same equation \refeq(non-1), 
$U_\l(t,s) = T(t)^{\ast} e^{-i(t-s)H_\l} T(s)$,  breaking the uniqueness. 
Note, however, that $\left. H_{C-1, 0}\right\vert_{C_0^\infty(\R^d)}$ 
is essentially selfadjoint if $C\geq 1$ and it generates a unique 
dymanics for \refeq(non-1). Incidentally, $H_C(t)$ with $\s>2$ and $C>0$ 
has purely discrete spectrum with super-exponentially decreasing eigenfunctions 
by virtue of \refeq(unit-eq) and, this shows that the 
similarity of the appearance or the spectral properties  
of the Hamiltonian does not guarantees the same for the dynamics. 
Notice also very different dynamics of the corresponding 
classical mechanical particles for $C<1$ and $C\geq 1$. 

In the example above, the break down of uniqueness happens only 
when $\s>2$ and the situation is very different if $\s\leq 2$. 
This is true in general and we have shown 
in \cite{Y-1} and \cite{Y-2} that, for the case $N=1$, 
if $A(t,x)$ and $\ph(t,x)$ are smooth and grow linearly 
or quadratically as $|x|\to \infty$ respectively then, Eqn. \refeq(1) 
generates a unitary propagator uniquely with the invariant 
subspace $\Si(2)$ when $V(t,x)$ is locally and spatially 
critically singular for the selfadjointness of $H(t)$ and 
spatial singularities of $\pa_t V(t,x)$ can be stronger 
than those of $V(t,x)$ itself. For example, it is proven that, when 
centers of forces $y_1(t), \dots, y_M(t)\in \R^3$ are moving 
smoothly, 
\bqn \lbeq(example)
i \pa_t u =  \left(-\frac1{2m}\lap  + \sum_{l=1}^M \frac{Z_l}
{|x_j-y_l(t)|^{\gamma}}\right)u 
\eqn 
generates a unique dynamics if $\gamma<3/2$ while 
$\pa_t |x-y_l(t)|^{-\gamma}$ are 
$-\lap$-form bounded only when $\gamma\leq 1$ (see \cite{FMS} 
for the result for $N$-body Coulomb system). 

Then, it is the purpose of this paper that we improve 
and extend the results of \cite{Y-1, Y-2} to $N$-particle 
systems, viz. by restricting the 
behavior of $A(t,x)$ and $\ph(t,x)$ as $|x|\to \infty$  
as above, we build a theory which guarantees the 
existence and the uniqueness of unitary propagators 
and which is general enough to cover most of conceivable 
applications in physics. We simultaneously show 
under a mild additional condition that the propagators 
thus obtained possess $\Si(2)$ as an invariant 
subspace with the properties mentioned above. 

We now state main results of this paper precisely. 
For a function $f(t,x, \dots)$ of $(t,x,\dots)$ 
and $l=0,1, \dots, \infty$, we write $f \in C^l_x$, 
$f \in C^l_{(t,x)}$ and etc. if $f$ is of class $C^l$ 
with respect to variables $x$, $(t,x)$ 
and etc.respectively. The skew symmetric $d\times d$ matrix 
\bqn \lbeq(4)
B(t,x)=(B_{jk}(t,x))= (\pa A_k/\pa x_j- \pa A_j/\pa x_k), 
\quad j,k=1, \dots, d
\eqn 
is the magnetic field generated by $A(t,x)$.  
Here in \refeq(4), $x=(x_1, \dots, x_d) \in \R^d$ and 
$x_j$ is not the position of $j$-th particle. We apologize 
for this double use of the notation and hope this causes no 
confusions. For a vector $a\in \R^n$ and an $n\times m$ matrix 
$C$, $|a|$ and $|C|$ are respectively the Euclidean length 
of $a$ and the norm of $C$ as a linear operator from $\C^m$ 
to $\C^n$ and $\la a \ra= (1+|a|^2)^{1/2}$. 
For multi-index $\a=(\a_1, \dots, \a_n)$, 
$|\a|=\a_1+ \cdots+ \a_n$. 

\bgass \lbass(A-1) Potentials $\ph(t,x)$ and 
$A(t,x)$ are real valued, $\ph, A \in C_x^\infty $ and, 
for any multi-index $\a$, 
$\pa_x^\a \ph \in C^0_{(t,x)}$ and 
$\pa_x^\a A\in C^1_{(t,x)}$. Moreover, 
the followings are satisfied for compact intervals 
$I \subset \R$: 
\ben 
\item[{\rm (1)}] For any $\a$ with $|\a|\geq 2$, 
there exists a constant $C_\a$ such that 
\bqn \lbeq(ph-1)
|\pa_x^\a \ph(t,x)|\leq C_\a, 
\quad (t,x)\in I \times \R^d.
\eqn 
\item[{\rm (2)}] For any $\a$ with $|\a|\geq 1$, 
there exist $\ep_\a>0$ and $C_\a$ such that
\begin{align} 
& |\pa_x^\a B(t,x)|\leq C_\a \ax^{-1-\ep_\a}, \lbeq(B-1) \\ 
& |\pa_x^\a A(t,x)|+  |\pa_x^\a \pa_t A(t,x)|
\leq C_\a, \quad (t,x)\in I \times \R^d. \lbeq(AB-1)
\end{align} 
\een 
\edass 
We remark that \refeq(B-1) implies 
$\lim_{|x|\to \infty}B(t,x)=B(t)$ exists uniformly 
for $t \in I$ and $|B(t,x)-B(t)|\leq C \ax^{-\ep}$, $\ep>0$. 
Thus, $B(t,x)$ is spatially a long range perturbation 
of a constant magnetic field $B(t)$. 

We assume that $V(t,\ux)$ is the sum of potentials 
$V_D(t,\ux_{D,r})$ of $|D|$-body interactions among particles 
in $D\subset (1, \dots, N)$ for the case $|D|\geq 2$
and  potentials $V_j(t,x_j)$ of single-body external forces 
acting on the $j$-th particle for the case $D=\{j\}$, $1\leq j \leq N$ 
(we may also consider that $V_j(t,x_j)$ is the non-smooth part of the 
electric scalar potential): 
\bqn 
V(t,\ux) = \sum_{D \subset (1, \dots, N)} 
V_D(t,\ux_{D,r}).
\lbeq(3) 
\eqn 
Here, if $|D|\geq 2$, $V_D(t,\ux_{D,r})$ are functions of 
$\ux_{D,r}$, the positions of particles in $D$ relative to 
the center of mass $x_{D,c}$ of $D$: 
\bqn \lbeq(cmass-j)
x_{D,c}=\sum_{j\in D} m_j x_j/ \sum_{i\in D} m_j 
\eqn  
and, we define $\ux_{D,r}=x_j$ as a convention . 
  
For stating the conditions on $V(t,\ux)$ precisely and 
also for later uses, we introduce some notation. We write 
$X=\R^{Nd}$ and define the inner product of 
$\ux=(x_1, \dots, x_N)$ and $\uy=(y_1, \dots, y_N)\in X$ 
by 
\bqn \lbeq(innerpro)
(\ux, \uy)_X=  \sum m_j (x_j,y_j)_{\R^d}.
\eqn 
The configuration space of particles in 
$D=\{j_1, \dots, j_n\}\subset \{1, \dots, N\}$ is   
\[
X_D= \{\ux_D=(x_{j_1}, \dots, x_{j_n}) 
\colon x_{j_k}\in \R^d, \ k=1, \dots, n \}= \R^{nd}  
\]
which is considered in the natural way as a subspace of $X$.
The configuration space of the center of mass of $D$ is 
defined by 
\[
X_{D,c}=\{\ux_D=(x_{j_1}, \dots, x_{j_n})\in X_D \colon 
x_{j_1}=\dots=x_{j_n}\} \simeq \R^d .
\]
The projection of $\ux_D=(x_{j_1}, \dots, x_{j_n})$ to 
$X_{D,c}$ is given by $\ux_{D,c}=(x_{D,c}, \dots, x_{D,c})$. 
The configuration space of the motion of particles in $D$ 
relative to $x_{D,c}$ is the orthogonal complement of 
$X_{D,c}$ within $X_D$:   
\bqn \lbeq(Dcc)
X_{D,r}= X_D \ominus X_{D,c} \simeq \R^{d(n-1)}, \quad 
\ux_D - \ux_{D,c}=(r_{j_1}, \dots, r_{j_n}). 
\eqn 
We take $x_{D,c}$ as the coordinates of $X_{D,c}$ and choose 
coordinates $x_{D,r}$ of $X_{D,r}$ (e.g. $x_{D,r}=x_2-x_1$ 
if $D=\{1,2\}$) such that 
\bqn \lbeq(measure)
dx_{D,c}dx_{D,r} = dx_D. 
\eqn
If $D=\{j\}$, we define 
$X_{D,r}=\R^d$, $X_{D,c}=\{0\}$ and $\ux_{D,r}=x_j$ 
as a convention.  
\[
n_D= \dim X_{D,r} = (|D|-1)d, \mbox{\ if} \ |D|\geq 2; \quad 
n_D= d , \mbox{\ if}\ |D|=1.  
\]
Recall that for Banach spaces $X_1, \dots, X_n $ 
which are subspaces of a linear topological space $Y$, the sum space 
$\Si=\sum X_j$ and intersection space $\lap= \cap X_j$ are 
Banach spaces with the respective norms 
\[
\|u\|_{\Si}= \inf\{\sum \|u_j\| \colon u=u_1+ \cdots+ u_n\}, 
\quad 
\|u\|_{\lap}= \|u\|_{X_1}+ \cdots+ \|u\|_{X_n} .
\]

\bgdf \lbdf(V-1)
For $D \subset \{1, \dots, N\}$, 
$1\leq a, p \leq \infty$ and 
compact intervals $I \subset \R$, we define three Banach spaces 
$\Ig^{a,p}_D(I)$ and $\tilde{\Ig}^{\infty, p}_{D}(I)$ by 
\begin{gather} \lbeq(V-1)
\Ig^{a,p}_{D}(I) 
= L^a(I, L^{p}(X_{D,r}))+ L^1(I, L^\infty(X_{D,r})). \\
\tilde{\Ig}^{\infty,p}_{D}(I) 
= C(I, L^{p}(X_{D,r}))+ L^1(I, L^\infty(X_{D,r})). \\
{\Ig}^{cont, p}_{D}(I) 
= C(I, L^{p}(X_{D,r}))+ C(I, L^\infty(X_{D,r})). 
\end{gather} 
For functions $V_D(t,\ux_D)$ of $(t,\ux_D)\in \R \times X_D$, 
we say $V_D \in \Ig^{a,p}_{\loc,D}$ 
(resp. $V_D \in \tilde{\Ig}^{\infty,p}_{\loc,D}$) if 
$V_D \in \Ig^{a,p}_{D}(I)$ 
(resp. $V_D \in \tilde{\Ig}^{\infty,p}_{\loc,D}(I)$) 
for compact intervals $I$.  Abusing notation, 
we write ${\Ig}^{cont, p}_{D}$ for 
${\Ig}^{cont, p}_{D}(\R)$. 
\eddf 

\noindent 
We have 
$\Ig^{a,p}_{D}(I)\subset \Ig^{\tilde{a},\tilde{p}}_{D}(I)$ 
if $\tilde{a}\leq a$ and $\tilde{p} \leq p$ and 
$\Ig^{cont,p}_D(I)\subset \tilde{\Ig}^{\infty,p}_{D}(I) \subset 
\Ig^{a,p}_{D}(I)$ for any $1\leq a \leq \infty$. 
We define $a(p_D)$ by 
\bqn \lbeq(index-1)
\frac1{a(p_D)}=1- \frac{n_D}{2p_D}, \ \ \mbox{for} \ \ 
\frac{n_D}2<p_D \leq \infty. 
\eqn 

\bgass \lbass(V-1) $V(t,\ux)$ is given by \refeq(3) with 
$V_D(t,\ux_{D,r})$ which satisfies either 
$V_D\in \tilde{\Ig}^{\infty, n_D/2}_{\loc, D}$ or 
$V_D \in \Ig^{a(p_D),p_D}_{\loc,D}$ for some 
$n_D/2 < p_D \leq \infty$.   
\edass 

\noindent 
As $a(p_D)$ decreases with $p_D$, $V_D\in \Ig^{a(p_D),p_D}_{\loc,D}$ is 
the smoother in $t$ if it is locally the more singular in $\ux_{D,r}$. 
$V_D$ is not necessary $-\lap_{\ux_{D,r}}$-bounded 
when $n_D \leq 4$.

We define $D^c=\{1, \dots, N\}\setminus D $ and  
$\Hg_D= L^2(X_{D,c}) \otimes L^2(X_{D^c})$. We have  
\[
\Hg= L^2(X_{D,r}) \otimes 
\Hg_D = L^2(X_{D,r}, \Hg_D), \quad D \subset 
\{1, \dots, N\}.
\]
Using the index $p_D$ of \refass(V-1) 
for $V_D(t,\ux_{D,r})$, we define $l_D$ and $\th_D$ by 
\bqn \lbeq(pair-1)
\frac1{l_D}=\frac1{2}-\frac{1}{2p_D}, \quad  
\frac1{\th_D} = \frac{n_D}{4p_D}. 
\eqn 
and, for intervals $I$, define Banach space $\Xg(I)$ 
of functions of $(t,\ux)\in I \times X$ by 
\begin{gather}
\Xg(I) = \cap_{D \subset\{1, \dots, N\}} 
L^{\th_D}(I, L^{l_D}(X_{D,r}, \Hg_D))\cap C(I, \Hg),
\lbeq(Xg)
\\
\lbeq(normXg)
\|u\|_{\Xg(I)} = \sum_{D}
\|u\|_{L^{\th_D}(I, L^{l_D}(X_{D,r}, \Hg_D))}+ \|u\|_{C(I, \Hg)}. 
\end{gather}
For $k=0,1, \dots$, we write $\Si(k)$ for 
$\Si(k)=\{u \colon x^\a \pa^\b u\in L^2(\R^n), \ 
|\a+\b|\leq k\}$ indiscriminately of the dimension of the 
space $\R^n$. $\Si(k)$ is the Hilbert space  
with the norm $\|u\|_{\Si(k)}$: 
\[
\|u\|_{\Si(k)}^2 =
\sum \left\{\|x^\a \pa^\b u\|_{L^2}^2 \colon 
|\a+\b|\leq k\right\}.
\]
$\Si(-k)$ is the dual space of $\Si(k)$ with respect to 
the inner product of $L^2(\R^n)$. 

\bgth \lbth(propa-1)
Suppose that $(\ph, A)$ and $V$ satisfy \refasss(A-1, V-1) 
respectively. Then, there uniquely exists a unitary propagator 
$\{U(t,s) \colon t,s \in \R\}$ 
for Eqn. \refeq(1) such that, 
for any $s\in \R$ and $f \in \Hg$, $u(t)= U(t,s)f$ 
satisfies the following properties: 
\ben 
\item[{\rm (1)}]  For compact intervals $I \subset \R$, 
$u \in \Xg(I)$ and for a constant $C$, 
\bqn \lbeq(hgxg-cont)
\|u\|_{\Xg(I)}\leq C\|f\|_{\Hg}, \quad f \in \Hg.
\eqn
\item[{\rm (2)}] The function $u(t)$ is a locally 
absolutely continuous 
(AC for short) function of $t\in \R$ with values in $\Si(-2)$  
and it satisfies in $\Si(-2)$ the equation 
\bqn \lbeq(S)
i\frac{du}{dt} =  H(t)u  , \quad {\rm a.e.}\,  
t\in \R.
\eqn 
\een
\edth 
\bgrm \lbrm(stri) 
The pair of indices $(\l, \s)$ is called 
$D$-admissible  Strichartz pair if it satisfies 
\bqn \lbeq(D-admissible)
0 \leq \frac2{\s}= 
n_D\left(\frac12- \frac1{\l}\right) \leq 1.
\eqn 
Hence, 
$(l_D,\th_D)$ of \refeq(pair-1) is a $D$-admissible pair. 
Interpolating \refeq(hgxg-cont) with with the unitary 
property of the propagator 
$\|u\|_{L^{\infty}(I, L^{2}(X_{D,r}, \Hg_D))}=\|f\|_{\Hg}$, we 
see that $u(t)=U(t,s)f$ satisfies the  
Strichartz inequality for all $D$-admissible Strichartz pairs 
$(\l,\s)$ and $(\m,\t)$ such that $2\leq \l,\m \leq l_D$:
\bqn 
\sup_{s\in I}\|U (t,s)f\|_{L^\s(I, L^{\l,2}_{D_1})} 
\leq C_I \|f\|_{\Hg} \lbeq(stri-f1) 
\eqn 
and, hence, the two others: 
\begin{gather} 
\sup_{s\in I} \left\| \int_{I} U(s,t)u(t) dt \right\|_{\Hg}
\leq C_I \|u\|_{L^{\s'}(I, L^{\l',2}_{D_1})},  
\lbeq(stri-f2) \\
\left\|\int_{s}^t U(t,r)u(r) dr 
\right\|_{L^{\s}([s,s+L], L^{\l,2}_{D_1})}
\leq C_L \|u\|_{L^{\t'}([s,s+L], L^{\m',2}_{D_2})}, 
\lbeq(stri-f3)
\end{gather}
where $\s',\l'$ and etc. are dual exponents of $\s, \l$ 
and etc: $1/\s+ 1/\s'=1$ and etc.(we refer to the proof of 
\reflm(stri) to see how \refeq(stri-f2) and \refeq(stri-f3) 
follow from \refeq(stri-f1)). 
Here we recall that $l_D=2p_D/(p_D-1)$, 
which implies the smaller set of $D$-admissible pairs 
for larger $p_D$. This 
seemingly contradicting phenomenon happens because we have chosen 
the smallest possible $a(p_D)$ for a given $p_D$ such that 
$V_D \in \Ig^{a, p_D}_D$ for obtaing the most general statement 
of the theorem. Thus, if $V_D \in \Ig^{a, p_D}_D$ is satisfied 
for $a>a(p_D)$, then $l_D$ can be replaced by the larger 
$2q_D/(q_D-1)$ 
such that  $a=a(q_D)$. 
In particular, if $V_D \in \tilde{\Ig}^{\infty,p}_{D}$ 
for $p\geq n_D/2$,  then, Strichartz estimates are 
satisfied for the full range of $\l$: $2\leq \l \leq 2n_D/(n_D-2)$. 
\edrm

For $n_D/2\leq p_D\leq \infty$ of \refth(propa-1), we define 
\begin{gather}
\tilde p_{D}= \max (2, p_D), \quad b_{D}= \frac{4p_D}{4p_D-n_D}, \lbeq(b-D) \\
q_{D}=\frac{2n_{D} p_D}{n_{D}+ 4p_{D}} 
\ \mbox{if}\ n_{D} \geq 4, \quad   
q_{D}=\frac{2p_D}{p_D+1} \lbeq(q-D)
\ \mbox{if}\ n_D =3.  
\end{gather}
\bgass \lbass(V-2) $V(t,\ux)$ is given by \refeq(3) with $V_D(t,\ux_D)$ 
such that 
$V_D \in \Ig^{cont, \tilde p_D}_D$ and $\pa_t V_D \in \Ig_{D,\loc}^{b_D,q_D}$ 
for some $n_D/2\leq p_D\leq \infty$.   
\edass

\bgrm Definition \refeq(q-D) may be written as 
\[
\frac1{q_D}=\frac1{p_D}+
\left(\frac{2}{n_D}-\frac1{2p_D}\right), \ n_D\geq 4; 
\quad  
\frac1{q_D}=\frac1{\tilde{p}_D}+ 
\left(\frac12+ \frac1{2p_D}-\frac1{\tilde{p}_D}\right),\ n_D=3. 
\]
Thus, as for the singularities of the type $|x|^{-a}$, 
$\pa_t V_D(t,x_{D,r})$ can be more singular than $V_D(t,x_{D,r})$ by  
$C|x_{D,r}|^{-\left(2-\frac{n_D}{2p_D}\right)+\ep}$, $\ep>0$ if $n_D\geq 4$ 
and, if $n_D=3$, by $C|x_{D,r}|^{-3\left(\frac{1}{2}-\frac1{2p_D}\right)+\ep}$ 
if $p_D\geq 2$ 
and $C|x_{D,r}|^{-\frac{3}{2p_D}+\ep}$ if $3/2\leq p_D\leq 2$. 
$\ep>0$. Thus, the exponents are $-1+\ep$ for all $n_D\geq 3$ if 
$p_D=n_D/2$ whereas for large $p_D$ they are close to $-2$ if 
$n_D\geq 4$ and to $-3/2$ if $n_D=3$.  
\edrm 

\bgth \lbth(2)
Suppose that $(\ph,A)$ and $V(t,\ux)$ 
satisfy \refasss(A-1, V-2) respectively.  Suppose, in addition, 
$\ph \in C^1_{(t,x)}$ and  
\bqn \lbeq(th-a2)
|A(t,x)|+ |\pa_t A(t,x)| \leq C \ax, \ \ 
|\pa_t \ph(t,x)|\leq C \ax^2, \ \ (t,x) \in I \times \R^d  
\eqn 
for compact intervals $I$. Then, 
$U(t,s)$ of \refth(propa-1) satisfies the following: If  $f\in \Si(2)$,   
then $u(t)=U(t,s)f\in C(\R, \Si(2)) \cap C^1(\R, \Hg)$ 
and $\pa_t u \in\Xg(I)$ for any compact interval $I$.
It satisfies Eqn. \refeq(1) as an evolution equation in $\Hg$. 
\edth 
\bgrm The proof of \refth(2) actually shows that, for 
$f \in \Si(2)$, the solution $u(t)$ satisfies 
$\pa_t u \in \Xg(I)$ for compact intervals $I$. 
Since $V u \in \Xg(I)$ if $u\in C(I,\Si(2))$ and 
$V$ satisfies \refass(V-2), we have $H_0(t)u\in \Xg(I)$ 
as well. Here $H_0(t)$ is the 
Hamiltonian for $N$ independent particles in the field: 
\bqn 
H_0(t)= \sum_{j=1}^N 
\left(\frac1{2m_j}(-i\hs\nabla_j - e_j A(t,x_j))^2 + 
e_j \ph(t,x_j)\right).   \lbeq(indep-H)
\eqn 
\edrm  

We describe here the plan of the paper. 
In Sec. 2 we 
record some results which will be used in later sections: 
We first recall some results of \cite{Fu, Y-2} on 
fundamental solutions, i.e. integral kernels of the 
unitary propagators of single particle Schr\"odinger 
equations and, prove Strichartz estimates of new type 
which is tailored for our purpose for 
the propagator associated with the Hamiltonian $H_0(t)$.   
With this new Strichartz estimates, 
we prove \refth(propa-1) in Sec. 3. The argument is 
based on the contraction mapping principle and is a 
straightforward extension of that in \cite{Y-1, Y-2}.  
We prove \refth(2) in Sec. 4 after a few preparations. 
In subsec. 4.1, we apply gauge transformation to 
Eqn. \refeq(1) and reduce the poof to the case when 
$\ph(t,x)\geq C \ax^2$ with a sufficiently large constant 
$C>0$.  Under this condition, we prove that $H_0(t)$ with 
domain $D(H_0(t))= \Si(2)$ is selfadjoint and $H_0(t)\geq 1$ 
for all $t \in \R$. We complete the proof of \refth(2) in 
subsec. 4.2. 

Most of the notation is standard. 
For Banach spaces $X$ and $Y$, $\Bb(X,Y)$ is the Banach 
space of all bounded operators from $X$ to $Y$ and 
$\Bb(X)=\Bb(X,X)$.  The Laplacian 
in Euclidean space of various dimensions is denoted  
indiscriminately by $\lap$. If $K(x,y)$ is a function of 
$(x, y) \in \R^d \times \R^d$, then 
$\frac{\pa^2 K}{\pa x \pa y}$ is the $d\times d$-matrix 
with $(j,k)$ elements $\frac{\pa^2 K}{\pa x_j \pa y_k }$, 
$1\leq j,k \leq d$.  For $n=0,1,\dots$, $O(\la \ux \ra^n)$ 
and etc. is a function which is bounded by 
$C\la \ux \ra^n$ and etc. Various constants are denoted by the 
same letter $C$ when their specific values are not important 
and the same $C$ may differ from one place to the other.    
{\it In what follows in this paper, 
we arbitrarily take and fix a (large) compact interval 
$I_0$ and assume that time variables and intervals are 
always inside $I_0$.} 

\section{Preliminaries}

In this section we record several known facts which we use 
for proving Theorems. 

\subsection{Fundamental solutions} 

We use the following well known theorem on single particle 
Schr\"odinger equations (\cite{Fu, Y-2}): 
\bgth \lbth(fund) Suppose that $A$ and $\ph$ satisfy 
\refass(A-1), $m>0$ and $e\in \R$. Then, there exists 
a unique unitary propagator  
$\{U_{sing}(t,s)\colon t,s\in \R\}$ on $L^2(\R^d)$ for 
the Schr\"odinger equation 
\bqn \lbeq(5)
i\pa_t u (t,x)= \left(\frac1{2m}(-i\nabla -e A(t,x))^2 + 
e\ph(t,x)\right) u (t,x). 
\eqn 
The propagator satisfies the following properties: 
\ben 
\item[{\rm (1)}] For $f \in \Si(k)$,  
$u(t)=U_{sing}(t,s)f$ satisfies 
$u\in C(\R,\Si(k)) \cap C^1(\R, \Si(k-2))$, $k=0,1,\dots$. 
In particular, $U_{sing}(t,s)$ is an isomorphism of 
$\Sg(\R^d)$. 
\item[{\rm (2)}] There exists $T>0$ such that, for $0<|t-s|<T$, 
$U_{sing}(t,s)$ is an oscillatory integral operator 
(OIOp for short) of the form  
\bqn \lbeq(d-prop)
U_{sing}(t,s)f(x) = \frac{m^{d/2}}
{(2\pi i(t-s))^{d/2}}\int_{\R^d} e^{iS(t,s,x,y)}
b(t,s,x,y)f(y)dy.
\eqn 
Here $S(t,s,x,y)$ and $b(t,s,x,y)$ satisfy 
the following properties:
\ben 
\item[{\rm (a)}] For any $\a, \b$, $S\in C_{(x,y)}^\infty$ 
and $\pa_x^\a \pa_y^\b S \in C^1_{(t,s,x,y)}$. 
For $|\a|+|\b|\geq 2$, there exists a constant $C_{\a\b}$ 
such that 
\bqn \lbeq(phase)
\left|
\pa_x^\a \pa_y^\b\left(S(t,s,x,y)- 
\frac{m(x-y)^2}{2(t-s)}\right)
\right|\leq C_{\a\b}, \ \  
(x,y)\in \R^d \times \R^d. 
\eqn 
\item[{\rm (b)}] For any $\a, \b$, $b\in C_{(x,y)}^\infty$ 
and $\pa_x^\a \pa_y^\b b \in C^1_{(t,s,x,y)}$. There exists a 
constant $C_{\a\b}$ such that  
\bqn \lbeq(amp)
|\pa^\a_x\pa^\b_y (b(t,s,x,y)-1)|\leq C_{\a\b}|t-s|, \quad  
(x,y) \in \R^d \times \R^d. 
\eqn 
\een
\een
\edth 
\noindent
Incidentally $S(t,s,x,y)$ is the action 
integral of the classical path $(p(\t), q(\t))$ 
corresponding to \refeq(5) such that $q(s)=y$ and 
$q(t)=x$, viz.  
\[
\frac{dq}{d\t}=\frac{\pa h}{\pa p}, \ \ 
\frac{dp}{d\t}=-\frac{\pa h}{\pa p}, \ \ 
h(\t,p,q)=\frac1{2m}(p -e A(\t,q))^2 + e \ph(\t,q).
\]
When $V=0$, $H(t)=H_0(t)$ and \refeq(1) becomes 
\begin{gather} \lbeq(6)
i\pa_t u = (H_{0,1}(t)+ \cdots+ H_{0,N}(t))u, \\
H_{0,j}(t)=\frac1{2m_j}(-i\nabla_j -e_j A(t,x_j))^2 + 
e_j\ph(t,x_j), \ \ j=1, \dots, N.  \lbeq(7)
\end{gather}
The unitary propagator for \refeq(6) is given by 
the tensor product:    
\bqn \lbeq(unpert)
U_0(t,s) = U_{0,1}(t,s) \otimes \cdots \otimes U_{0,N}(t,s)
\ \mbox{on}\ \Hg=\otimes_{j=1}^N L^2(\R^d) 
\eqn 
of the propagators $U_{0,j}(t,s)$ for the $j$-th particle:
\bqn 
i\pa_t u = H_{0,j}(t)u, \quad j=1, \dots, N. 
\eqn
By virtue of \refth(fund), 
$\{U_0(t,s)\colon -\infty<t,s<\infty\}$ 
is strongly continuous in $\Si(k)$ and $C^1$ from 
$\Si(k)$ to $\Si(k-2)$ for any $k=0,1, \dots$ and, 
it is strongly $C^1$ in $\Sg(\R^{Nd})$. There exists $T>0$ 
such that for $0<|t-s|<T$  all $U_{0,j}(t,s)$ are OIOp's 
of the form    
\bqn \lbeq(d-propj)
U_{0,j}(t,s)f(x) = \frac{m_j^{d/2}}
{(2\pi i(t-s))^{d/2}}\int_{\R^d} e^{iS_j(t,s,x,y)}
b_j(t,s,x,y)f(y)dy, 
\eqn 
with $S_j(t,s,x,y)$ and $b_j(t,s,x,y)$ which satisfy the 
properties corresponding to (a) and (b) of \refth(fund). 
{\it We take this $T>0$ and, in what follows, we will 
make it further smaller when it becomes necessary.} 

For $D =\{j_1, \dots, j_n\} \subset \{1, \dots, N\}$, 
$U_{0,D}(t,s)$ is the unitary propagator on $L^2(X_D)$ 
for $n$ independent particles inside $D$:  
\bqn \lbeq(ud)
U_{0,D}(t,s)= U_{0,j_1}(t,s) \otimes \cdots 
\otimes U_{0,j_n}(t,s).
\eqn 
We often consider $U_{0,D}(t,s)$  as an operator on 
$L^2(X)= L^2(X_D) \otimes L^2(X_{D^c})$ by 
identifying it with 
$U_{0,D}(t,s) \otimes {\bf 1}_{L^2(X_{D^c})}$, 
where ${\bf 1}_{L^2(X_{D^c})}$ is the identity operator of 
$L^2(X_{D^c})$. 

\subsection{Strichartz estimates} 
From the decomposition  
$X=X_{D^c} \oplus X_{D,c} \oplus X_{D,r}$, 
we have $\Hg = L^2(X_{D,r}, \Hg_D)$, $\Hg_D=L^2(X_{D^c} \oplus X_{D,c})$, 
see \refeq(measure).   
Then, for $1\leq p, q \leq \infty$, we define 
\begin{gather}
L^{p,q}_D(X)= L^p(X_{D,r}, L^q(X_{D^c}\oplus X_{D,c})), \\
\|u\|_{L^{p,q}_D} = \left(\int_{X_{D,r}} 
\|u(\ux_{D,r}, \cdot)\|_{L^q(X_{D^c}\oplus X_{D,c})}^p d\ux_{D,r} \right)^{1/p}.
\end{gather}
Recall $n_D=\dim X_{D,r}=(|D|-1)d$ if $|D|\geq 2$ and 
$n_D=d$ if $|D|=1$. The following lemma is the extension 
to $N$ independent particles of 
the well-known $L^p$--$L^q$ estimates for single particle  
Schr\"odinger equations \refeq(5).

\bglm \lblm(lpq)
Let $1\leq q\leq 2 \leq p \leq \infty$ satisfy 
$1/p+ 1/q =1$. Then, for any $D \subset \{1, \dots, N\}$, 
there exists a constant $C$ such that, for 
$0<|t-s|\leq T$, 
\bqn \lbeq(p-q)
\|U_0(t,s) u \|_{L^{p,2}_D(X)}\leq 
C |t-s|^{-n_D(1/2-1/p)}
\|u\|_{L^{q,2}_D(X)}.
\eqn 
\edlm 
\bgpf We prove the lemma when $D=\{1, \dots, N\}$,  
omitting subscript $D$ from various notation, e.g.  
$X_c=X_{D,c}$, $X_r= X_{D,r}$ and 
$\ux=(x_c, \ux_r) \in X=X_c \oplus X_r$. The proof for 
other cases is similar. From \refeq(unpert) and 
\refeq(d-propj) 
\begin{multline} 
U_0(t,s) u(x_c, \ux_r) =   
\frac{(m_1 \cdots m_N)^{d/2}}{(2\pi i (t-s))^{dN/2}} \\
\times \int_{\R^{Nd}} e^{i\sum S_j(t,s,x_j,y_j)}
\prod b_j(t,s,x_j, y_j)u(y_c, \uy_r)dy_c 
d\uy_r. 
\end{multline}
Using the notation \refeq(Dcc) for 
$\ux= \ux_c+ (r_1, \dots, r_N)$ and the corresponding  
for $\uy=\uy_c+ (s_1, \dots,  s_N)$, we write   
\begin{gather}
F(t,s,\ux,\uy)\equiv  \sum_{j=1}^N S_j(t,s,x_j,y_j) 
= \sum_{j=1}^N S_j(t,s, x_c+r_j, y_c+ s_j), \\  
B(t,s,\ux,\uy)\equiv \prod_{j=1}^N b_j(t,s,x_j, y_j)= 
\prod_{j=1}^N b_j(t,s,x_c+ r_j, y_c+ s_j). 
\end{gather}
Then, \refeq(phase) and \refeq(amp) respectively imply 
\begin{gather}
\frac{\pa^2 F}{\pa{x_c} \pa{y_c}}
=\sum_{j=1}^N 
\frac{\pa^2 S_j}{\pa x_c \pa{y_c}}(t,s, x_c+r_j, y_c+ s_j)
= \sum_{j=1}^N \frac{m_j}{t-s}{\bf 1}_d + O(1), 
 \lbeq(F) \\
|\pa_{x_c}^\a \pa_{y_c}^\b B(t,s,\ux,\uy)|\leq C_{\a\b}, 
\quad |\a|+|\b|\geq 0 , 
\end{gather}
where ${\bf 1}_d$ is the $d\times d$ unit matrix and 
$O(1)$ is $d\times d$ matrix whose components are functions 
bounded along with all derivatives with respect to the spatial 
variables $(\ux, \uy) \in \R^{Nd} \times \R^{Nd}$. 
Then, the Minkowski inequality and the Asada-Fujiwara 
$L^2$-boundedness theorem for OIOp's (\cite{AF}) yield    
\begin{align*}
& \frac{1}{(2\pi|t-s|)^{Nd/2}}\left\|\int_{X_r}
\left(\int_{X_c} e^{iF(t,s,\ux,\uy)}B(t,s,\ux,\uy)
u(y_c, \uy_r)dy_c \right)
d\uy_r\right\|_{L^2(X_c)} \\
& \qquad 
\leq \frac{1}{(2\pi|t-s|)^{Nd/2}}
\int_{X_r}\left\|\int_{X_c} e^{iF(t,s,x,y)}B(t,s,x,y)
u(y_c, \uy_r)dy_c \right\|_{L^2(X_c)} d\uy_r \\
& \qquad \leq 
\frac{C}{|t-s|^{d(N-1)/2}} 
\int_{X_r}\left\|u(\cdot, \uy_r)\right\|_{L^2(X_c)} d\uy_r.
\end{align*}
It follows that 
\bqn \lbeq(1-infty)
\|U_0(t,s) u(x_c, \ux_r)\|_{L^{\infty,2}_D}
\leq C|t-s|^{-d(N-1)/2} 
\|u(y_c, \uy_r)\|_{L^{1,2}_D} 
\eqn 
Thus, interpolating the inequality \refeq(1-infty) with 
the unitary property of $U_0(t,s)$: 
\[
\|U_0(t,s)u\|_{L^{2,2}_D} = \|u\|_{L^{2,2}_D},
\]
we obtain the desired result. 
\edpf

Recall that 
the pair of indices 
$(\l, \s)$ is called $D$-admissible if it satisfies 
\refeq(D-admissible).  We have the following 
set of Strichartz' estimates for $N$ independent 
particles in the external field (cf. \refrm(stri)). 

\bglm  \lblm(stri)
Let $D_1, D_2 \subset \{1, \dots, N\}$,  
$(\l, \s)$ and $(\m, \t)$ be admissible 
pairs for $D_1$ and $D_2$, respectively and 
$\l'$ and etc. be dual exponents of $\l$ 
and etc., viz. $1/\l+ 1/\l'=1$ and etc. 
Then, there exist a constant $C$ such that 
the following estimates are satisfied 
for intervals $I \subset I_0$ and 
$[s,s+L]\subset I_0$: 
\begin{gather} 
\|U_0 (t,s)f\|_{L^\s(I, L^{\l,2}_{D_1})} 
\leq C \|f\|_{\Hg},\quad s\in \R. \lbeq(stri-1) \\
\left\| \int_{I} U_0(s,t)u(t) dt \right\|_{\Hg}
\leq C \|u\|_{L^{\s'}(I, L^{\l',2}_{D_1})}, \quad 
s\in \R. \lbeq(stri-2) \\
\left\|\int_{s}^t U_0(t,r)u(r) dr 
\right\|_{L^{\s}([s,s+L], L^{\l,2}_{D_1})}
\leq C \|u\|_{L^{\t'}([s,s+L], L^{\m',2}_{D_2})}, 
\quad s\in I.  
\lbeq(stri-3)
\end{gather}
The estimate \refeq(stri-3) likewise holds if $[s,s+L]$ is 
replaced by $[s-L,s] \subset I_0$.  
\edlm  
\bgpf Define, for every fixed $s\in \R$,  
\[
U_{0,s}(t)= \left\{
\br{ll} U_0 (t,s), \quad & s-T <t< s+T, \\
0,  &  T \leq |t-s|. 
\er 
\right. 
\]
Then, $\{U_{0,s}(t)\colon t \in \R\}$ 
satisfies 
\[
\mbox{either $U_{0,s}(t)U_{0,s}(r)^\ast = U_0(t,r)$ or 
$U_{0,s}(t)U_{0,s}(r)^\ast = 0$}
\]
and \reflm(lpq) implies  
\begin{gather} 
\|U_{0,s}(t)f\|_{L_D^{2,2}}
\leq C \|f\|_{L^{2,2}_D}, \quad 
f \in L^{2,2}_D. \lbeq(st-98) \\
\|U_{0,s}(t)U_{0,s} (r)^\ast f\|_{L^{\infty,2}_D} 
\leq C |t-r|^{-n_D/2}\|f\|_{L^{1,2}_D}, \quad 
f \in L^{1,2}_{D}.  \lbeq(st-99)
\end{gather}
Thus, if we consider $\{U_{0,s}(t)\colon t \in \R\}$ 
as the family of operators 
acting on functions of $\ux_r \in X_{D,r}$ 
with values in the Hilbert space $\Hg_D=L^2(X_{D,c}\oplus X_{D,c})$, 
it satisfies the Keel-Tao conditions (\cite{KT}) for Strichartz 
estimates. Then, thanks to the fact that, 
for any Hilbert space $\Kg$,  
\bqn \lbeq(d)
L^{\th}(\R^{d_1}, L^{p}(\R^{d_2}, \Kg))^\ast 
= L^{\th'}(\R^{d_1}, L^{p'}(\R^{d_2},\Kg)) 
\eqn 
when $1\leq \th, \th'<\infty$ and  
$1<p,p'<\infty$ satisfy $1/\th+ 1/\th'=1$ and 
$1/p+1/p'=1$ (cf. e.g. \cite{DU}, pp. 97--100), 
the proof of Strichartz estimates presented in 
\cite{Sogge} can be applied almost word by word 
to the vector valued functions and produces 
\refeq(stri-1), \refeq(stri-2) and \refeq(stri-3) 
when $|I|<T$ and $L<T$ respectively. 

We next prove \refeq(stri-1), \refeq(stri-2) and \refeq(stri-3) 
when the time intervals $I=[T_1, T_2] \subset I_0$ and 
$[s, s+L] \subset I_0$ are of arbitrary size. Then we take a decomposition 
$t_0=T_1<t_1< \dots <t_n= T_2$ in such a way that 
$T/2< t_j - t_{j-1}<T$ for $j=1, \dots, n$. 
Then, the result \refeq(stri-1) for $|I|<T$ implies  
\begin{align*}
& \int_{T_1}^{T_2} \|U_0(t,s)f\|_{L^{\l,2}_{D_1}}^{\s} dt 
 = \sum_{j=1}^n \int_{t_{j-1}}^{t_j} 
\|U_0(t,s)f\|_{L^{\l,2}_{D_1}}^{\s} dt \\
& \hspace{2cm} = 
\sum_{j=1}^n \int_{t_{j-1}}^{t_j}
\|U_{t_{j-1}}(t)U_0(t_{j-1},s)f\|_{L^{\l,2}_{D_1}}^{\s} dt \\
& \hspace{2cm}\leq C \sum_{j=1}^n \|U_0(t_{j-1},s)f\|_{\Hg}^{\s}
\leq C n \|f\|_{\Hg}^{\s} 
\end{align*}
and \refeq(stri-1) for $[T_1, T_2]$ follows. The estimate 
\refeq(stri-2) follows from \refeq(stri-1) by the well 
known duality argument. For proving \refeq(stri-3), 
for shorting formulas, we 
write $F(t)=\int_s^t U_0(t,r)u(r) dr$ . We again 
decompose $[s,s+L]$: $t_0=s<t_1< \dots <t_n= s+L$ 
in such a way that $T/2< t_j - t_{j-1}<T$ for 
$j=1, \dots, n$. Then, denoting $\lap_j=(t_{j-1}, t_j)$, 
we have   
\begin{multline*}
\left\|F \right\|_{L^\s([s,L+s], L^{\l,2}_{D_1})}
\leq \sum_{j=1}^n 
\left\|F \right\|_{L^\s(\lap_j, L^{\l,2}_{D_1})} \\
= 
\sum_{j=1}^n 
\left\| 
U_{t_{j-1}}(t)
\int_s^{t_{j-1}} U_0(t_{j-1},r)u(r) dr
+ \int^t_{t_{j-1}} U_0(t,r)u(r) dr
\right\|_{L^\s(\lap_j, L^{\l,2}_{D_1})}.
\end{multline*}
We apply \refeq(stri-1) and \refeq(stri-2) 
to the first term in the sign of the norm 
and the short time result \refeq(stri-3) to 
the second. Then, we see that the right hand side is 
bounded by 
\begin{align*}
& \sum_{j=1}^n \left(
\left\| 
\int_s^{t_{j-1}} 
U_0(t_{j-1},r)u(r) dr
\right\|_{\Hg}
+ \|u\|_{L^{\t'}(\lap_j, L^{\m',2}_{D_2})} \right) \\
& \quad \leq C 
\sum_{j=1}^n 
\left( 
\|u\|_{L^{\t'}([s,t_{j-1}], L^{\m',2}_{D_2})} 
+ \|u\|_{L^{\t'}(\lap_j, L^{\m',2}_{D_2})} \right)\\
& \quad \leq C(n+1)^{1-1/\t'} 
\|u\|_{L^{\t'}([s,L+s], L^{\m',2}_{D_2})}.
\end{align*}
The proof of \refeq(stri-3) with $[s-L,s]$ is similar. 
This completes the proof. 
\edpf 

We use the following version of Christ-Kiselev lemma \cite{CK} 
which appears in \cite{Sogge} and which is also used in 
the proof of \reflm(stri). 
\bglm \lblm(ChKi) 
Let $Y$ and $Z$ be Banach spaces and assume that $K(t,s)$ 
is a strongly measurable function of $t,s \in (a,b)$, 
$-\infty \leq a <b \leq \infty$ taking values in $\Bb(Y,Z)$ 
such that $\|K(t,s)\|_{\Bb(X,Y)}$ is locally integrable. Set 
\[
Tf(t)= \int_a^b K(t,s) f(s) ds.
\]
Assume that 
\[
\|Tf\|_{L^q((a,b), Z)}\leq C_0 \|f\|_{L^p((a,b), Y)}
\]
Define 
\[
Wf(t) = \int^t_a K(t,s) f(s) ds 
\]
Then, if $1\leq p <q \leq \infty$, 
\[
\|Wf \|_{L^q((a,b), Z)}\leq C_0 C_{p,q} \|f\|_{L^p((a,b), Y)} 
\]
\edlm 

We often use the following estimates in the proof of 
\reflm(Ggs)
\bglm \lblm(L1-stri) 
Let $I=[s, s+L]$ or $I=[s-L, s]$. Then, there  exists 
a constant $C$ such that 
\bqn \lbeq(L1-stri)
\left\|\int_s^t U_0(t,r) u(r) dr \right\|_{\Xg(I)}
\leq C \|u\|_{L^1(I,\Hg)}.
\eqn 
\edlm 
\bgpf We prove the case  $I=[s, s+L]$. The other case may be proves 
similarly. Let $M=s+L$. Then 
\[
\left\|\int_s^M U_0(s,r) u(r) dr \right\|_{\Hg}\leq 
L \|u\|_{L^1(I,\Hg)}
\]
It follows by \refeq(stri-1) that 
\[
\left\|\int_s^M U_0(t,r) u(r) dr \right\|_{\Xg(I)}\leq 
C\|u\|_{L^1(I,\Hg)}
\]
Then, \refeq(L1-stri) follows by virtue of \reflm(ChKi). 
\edpf

\section{Proof of \refthb(propa-1)} 
The proof of \refth(propa-1) is the adaptation of 
that of Theorem 1 of \cite{Y-2}, using new function 
spaces $\Xg(I)$ defined above and 
new Strichartz estimates of \reflm(stri). 

Let $\Vg$ be the multiplication by $V(t,x)$: 
\[
(\Vg u)(t,\ux) = V(t,\ux) u (t,\ux)= 
\sum_{D \subset (1, \dots, N)} 
V_D(t,\ux_{D,r}) u (t,\ux)
\]
and $\Gg_s$ be the integral operator defined by  
\bqn \lbeq(gsdef)
(\Gg_s u)(t)= -i \int_s^t U_0(t,r)u(r) dr,
\eqn 
where $U_0(t,s)$ is the unitary propagator for $H_0(t)$ 
defined by \refeq(unpert).  
The Duhamel formula implies that \refeq(1) with 
the initial condition 
$u(s)=f\in \Hg$ is equivalent to the integral equation 
\bqn \lbeq(int-eq)
u(t) = u_0(t) + (\Gg_s \Vg u)(t),
\eqn 
where we wrote $u_0(t) =U_0(t,s) f$.
Define for intervals $I\subset I_0$ the function space 
$\Xg(I)$ by \refeq(Xg) by 
using indices $l_D$ and $\th_D$ of \refeq(pair-1). 
As was remarked previously $(l_D, \th_D)$ is a  
$D$-admissible pair. Along with $\Xg(I)$, we define 
another Banach space $\Xg^\ast (I)$ by 
\begin{gather*}
\Xg^\ast (I) = \sum_{D} 
L^{\th_D'}(I, L^{l',2}_D )+ L^1 (I, \Hg), \\
\|u\|_{\Xg^\ast(I)}= 
\inf \{
\sum_{D}
\|u_D\|_{L^{\th_D'}(I, L^{l_D',2}_D)}
+ \|u_1\|_{L^1 (I,\Hg)} \colon u = \sum_{D} u_D + u_1 \}, 
\end{gather*}
where $\th_D'$ and $l_D'$ are dual exponents of 
$\th_D$ and $l_D$ respectively. The space $\Xg^\ast(I)$ is almost 
the dual space of $\Xg(I)$ but not exactly. 

\bglm  For a constant $C$ independent of 
$I \subset I_0$ and $s\in I$ following 
statements are satisfied:  
\ben 
\item[{\rm (1)}] For $f \in \Hg$, $U_0(t,s) f \in \Xg(I)$.  
\item[{\rm (2)}] The multiplication $\Vg$ is 
bounded from $\Xg(I)$ to $\Xg^\ast (I)$ and 
\bqn 
\|\Vg u\|_{\Xg^\ast(I)}\leq C 
\max_{D} \|V_D\|_{\Ig_D^{a(p_D), p_D}(I)} \|u\|_{\Xg(I)}, \lbeq(est-V)   
\eqn 
\item[{\rm (3)}] Integral operator 
$\Gg_s$ is bounded from $\Xg^\ast (I)$ to $\Xg(I)$ and  
\bqn \lbeq(est-Ga)
\|\Gg_s u \|_{\Xg(I)} \leq C \|u\|_{\Xg^\ast(I)}.
\eqn 
\een 
\edlm 
\bgpf Write $a(p_D)=a_D$. 
Statement (1) is a result of \refeq(stri-1) and 
\refeq(stri-2). We have 
\[
\frac1{l_D}+ \frac1{p_D}= \frac1{l_D'} \ \ 
\mbox{and} \ \ 
\frac1{\th_D}+ \frac1{a_D}= \frac1{\th_D'}
\]
by the definition \refeq(index-1) and \refeq(pair-1). 
Thus, H\"older's inequality implies 
for $V_D= V_D^{(1)}+ V_D^{(2)}\in 
L^{l_D}(I, L^{p_D}(X_{D,r})+ L^{1}(I, L^{\infty}(X_{D,r})$ that 
\begin{gather*}
\|V_{D}^{(1)}u\|_{L^{\th_D'}(I, L^{l'_D,2}_D)}
\leq \|V_D^{(1)}\|_{L^{a_D}(I, L^{p_D}(X_{D,r}))} 
\|u\|_{L^{\th_D}(I, L^{l_D,2}_D)}, \\
\|V_{D}^{(2)}u\|_{L^{1}(I, \Hg)}
\leq \|V_D^{(2)}\|_{L^{1}(I, L^{\infty}(X_{D,r})} 
\|u\|_{L^{\infty}(I, \Hg)}. 
\end{gather*}
This implies \refeq(est-V) and statement (2) is proved. 
By virtue of \refeq(stri-3), $\Gg_s$ is bounded 
from the sum space 
$\Si=\sum_{D}L^{\th_D'}(I, L^{l_D',2}_D )$ to  
the intersection space 
$\cap_{D} L^{\th_D}(I, L^{l_D,2}_D )$. It is also 
bounded from $\Si$ to $C(I, \Hg)$, for $\Gg_s$ is 
bounded from $\Si$ into $L^{\infty}(I, \Hg)$, 
$\Gg_s u\in C(I, \Hg)$ if $u \in C(I, \Si(2))$
and $C(I, \Si(2))$ is dense in $\Si$. By virtue of 
Minkowski's inequality and \refeq(stri-1),  
$\Gg_s$ is bounded from $L^1(I, \Hg)$ to $\Xg(I)$. 
Thus, statement (3) is satisfied. . 
\edpf 

\paragraph{Proof of \refthb(propa-1)} 
Let $I \subset I_0$ be an interval. 
Then estimates \refeq(est-V) and \refeq(est-Ga) imply 
\[
\| \Gg_s \Vg u \|_{\Xg(I)}
\leq C \sum_{D} \|V_D\|_{\Ig^{a(p_D),p_D}_D(I)}\|u\|_{\Xg(I)}, 
\]
where $C$ is independent of $I$ and $s\in I$. 
It is obvious that 
$\|V_D\|_{\Ig^{a(p_D),p_D}_D(I)}\to 0$ as $|I|\to 0$ if $a(p_D)<\infty$ 
or $p_D<n_D/2$. This is also true when $p_D=n_D/2$ and 
$V_D \in \tilde{\Ig}^{\infty, n_D/2}(I)$, for 
$f_M(t)\equiv \|V_D(t,x_{D,r})\chi_{\{|V_D(t,x_{D,r})|>M\}}\|_{L^p(X_{D,r})}$ 
is continuous, $\lim_{M\to \infty}f_M(t)=0$ 
decreasingly, hence uniformly on $I$ by Dini's theorem and   
$\|V_D(t,x_{D,r})\chi_{\{|V_D(t,x_{D,r})|\leq M\}}\|_{L^{1,\infty}_D(I)} \leq 
M |I| \to 0$ as $|I|\to 0 $. Thus, $\Gg_s \Vg\colon \Xg(I) \to \Xg(I)$ is 
a contraction if $I$ is sufficiently small, and 
\refeq(int-eq) has a unique solution $u \in \Xg(I)$ 
for any $f \in \Hg$. It can be expressed as 
\[
u(t)=\Ga_t (1-\Gg_s\Vg)^{-1} T_s f, \quad 
(T_s f)(t) \equiv U_0(t,s)f, 
\]
where $\Ga_t$ is the evaluation operator at $t$, i.e.  
$\Ga_t w = w(t)$ for $w\in \Xg(I)$.  We define 
the operator $U(t,s)$ for $t,s \in I$ by 
\bqn 
U(t,s) = \Ga_t (1-\Gg_s\Vg)^{-1} T_s. 
\eqn
The proof of Theorem 1 of \cite{Y-2} shows that 
$\{U(t,s) \colon t,s\in I\}$ is a strongly continuous 
family of unitary operators in $\Hg$ 
and it satisfies Eqn. \refeq(CK) whenever $t,s,r\in I$. 
It is then well known (see e.g. \cite{K-3}) that such a family 
can be patched together to produce a globally defined strongly 
continuous family of unitary operators 
$\{U(t,s) \colon t,s\in \R\}$ 
in $\Hg$ which satisfies \refeq(CK). Then, property (1) of 
\refth(propa-1) is evidently satisfied.  

We prove (2). It suffices to prove it when 
$t,s\in I$ for sufficiently small intervals $I$ and   
$u$ satisfies \refeq(int-eq). By virtue of \refth(fund) and 
\refeq(unpert), $u_0(t)=U_0(t,s)f\in C^1(\R, \Si(-2))$ and 
$i\dot u_0(t)=H_0(t)u_0(t)$. Sobolev embedding theorem 
implies 
\[
\mbox{$\Si(2) \subset \cap_D L^{l_D,2}_D(X)$ and, hence, 
$\Si(-2) \supset \sum_D L^{l'_D,2}_D(X)$}
\]
by duality. 
Thus, $\Vg u \in \Xg^\ast(I) \subset L^1(I, \Si(-2))$ 
and the function 
\[
g(t)= \int^t_s U_0(s,r) V(r) u(r) dr 
\]
is locally AC with values in $\Si(-2)$ and is simultaneously 
continuous with values in $\Hg$ by \refeq(stri-2).
It follows that   
\[
u(t)= u_0(t)+ (\Gg_s\Vg u)(t)= u_0(t)-i U_0(t,s)g(t), 
\]
is $\Si(-2)$-valued locally AC and 
$i\dot u(t)= H_0(t)u(t) + V(t)u(t)$, a.e. $t$. 

Finally, we show that any $u \in \Xg(I)$  
which is locally AC with values in $\Si(-2)$ 
and which satisfies \refeq(1) with $u(s)=f\in \Hg$, 
$s\in I$ must do \refeq(int-eq) and, hence, is unique. 
To see this, we first note that 
$(H_0(s)u, v)= (u, H_0(s)v)$ for $u\in \Hg$ 
and $v \in \Si(2)$ and that, for $w\in \Si(-2)$
and $\ph\in \Si(2)$, $(w, U_0(s,t)\ph)= (U_0(t,s)w, \ph)$. 
This can be seen by approximating $u$ and $w$ respectively 
by sequences 
$u_n\in \Si(2)$ and $\w_n\in \Si(2)$ such that 
$\|u_n- u\|_\Hg \to 0$ and 
$\|w_n -w\|_{\Si(-2)}\to 0$ as $n\to \infty$. 
Then, for $\ph \in \Si(2)$ we have     
\begin{multline}
i\frac{d}{ds}(u(s), U_0 (s,t)\ph) 
=(H_0(s)u(s), U_0(s,t)\ph) + (V(s)u(s), U_0(s,t)\ph) 
\\ -(u(s), H_0(s)U_0(s,t)\ph) =(U_0(t,s)V(s)u(s), \ph), 
\quad 
{\rm a.e.}\, s.  
\end{multline}
It follows by integration that 
\[
(u(t), \ph) -(U_0(t,0)u(0), \ph)
= -i \int_0^t (U_0(t,s)V(s)u(s), \ph)ds 
\]
and $u(t)$ has to satisfy \refeq(int-eq). 
This completes the proof. 
\qed 

\section{Proof of \refthb(2)}

We begin by recording several preliminaries for the proof. 
We assume in what follows that $(\ph, A)$ satisfies the 
conditions of \refth(2).  

\subsection{Gauge transformation}
We define time dependent gauge transformation 
\[
T(t) u (\ux) = e^{itC \la \ux \ra^2}u(\ux), \quad 
(\Tg u)(t,\ux) = T(t) u(\ux), \quad t \in \R. 
\]

The following lemma is obvious. 

\bglm \lblm(gauge-trans)
\ben 
\item[{\rm (1)}] 
$\{T(t)\}$ is a strongly 
continuous unitary group in $\Hg$ 
\item[{\rm (2)}] For any interval $I$, 
$\Tg$ is an isomorphism of the Banach space $\Xg(I)$ and 
at the same time of the space $C(I, \Si(2))\cap C^1(I, L^2)$. 
\item[{\rm (3)}] $u(t,\ux)$ satisfies \refeq(1) 
if and only if $v(t,\ux)= (\Tg u)(t,\ux)$ does the same 
with $\tilde A(t,x)= A(t,x) -2tCx$ and $\tilde \ph(t,x) 
=\ph(t,x) + C\ax^2$ replacing $A$ and $\ph$ respectively.
New potentials $\tilde A(t,x)$ and $\tilde \ph(t,x)$ 
satisfy the conditions of \refth(2). 
\een
\edlm 

It follows from \reflm(gauge-trans) (3) that 
\begin{gather} 
i\pa_t u =  \tilde H_0(t)u + V(t,x) u \lbeq(1-a) \\
\tilde H_0(t)= -\sum_{j=1}^N 
\left(\frac1{2m_j}(\nabla_j - i e_j \tilde A(t,x_j))^2 + 
e_j \tilde \ph(t,x_j)\right)u,  \lbeq(2-a)
\end{gather} 
generates a unique unitary propagator 
which satisfies the properties of \refth(propa-1), 
which we denote by $\tilde U(t,s)$. Then, the 
uniqueness result of the theorem implies   
\[
U(t,s) = T(t)\tilde U(t,s)T(s)^{-1}
\]
and we may prove \refth(2) additionally assuming 
$\ph(t,x) \geq C\ax^2$,  which we do in what follows. 
The merit of doing so is that $H_0(t)$ will then become 
selfadjoint with common domain $\Si(2)$ and with a core 
$C_0^\infty(X)$, see below. 

For the proof of the next lemma we use the following 
well-known results on the $n$-dimensional 
quantum harmonic oscillator, $n=1,2, \dots$: 
\[
H_{os}=-\frac12 \lap + \frac12 x^2 \quad x\in \R^n.
\]
\begin{enumerate}
\item[{\rm (a)}] $H_{os}$ with domain 
$D(H_{os})=\Si(2)$ is selfadjoint in $L^2(\R^n)$, 
$H_{os}\geq n/2$ and 
$C_0^\infty(\R^n)$ is a core. 
\item[{\rm (b)}] For any set of 
multi-indices $\a,\b,\c$ and $\d$ with 
$|\a+\b+\c+\d|\leq 2$, the operator 
$x^\a \pa^\b H_{os}^{-1} x^\c \pa^\d$ has a 
bounded extension in $\Hg$. 
\item[{\rm (c)}] The integral kernel $G(x,y)$ of 
$H_{os}^{-1}$ satisfies for constants $C,c>0$, 
\bqn 
0< G(x,y) \leq \frac{C e^{-c|x-y|(1+|x|+|y|)}}{|x-y|^{n-2}}, 
\quad x,y \in \R^n.
\eqn
\end{enumerate}

We write $\pa_t u= \dot u$ and etc. hereafter. 

\bglm \lblm(res-h) 
Suppose that $A$ and $\ph$ satisfy the assumption 
of \refth(2) and that $\ph(t, x)\geq \frac12 \ax^2$, 
$(t,x) \in I_0 \times \R^d$. Then: 
\ben 
\item[{\rm (1)}] The operator $H_0(t)$ with domain 
$D(H_0(t))=\Si(2)$ is 
selfadjoint in $\Hg$, $H_0(t) \geq (Nd+1)/2$ 
and $C_0^\infty(X)$ is a core. 
\item[{\rm (2)}] $H_0(t)$ and $H_0(t)^{-1}$ are strongly 
differentiable functions of $t\in I_0$ with values in 
$\Bb(\Si(2), \Hg)$ and $\Bb(\Hg, \Si(2))$ respectively. 
\een
\edlm
\bgpf 
We may assume $e_j=m_j=1$, $1\leq j \leq N$. For proving (1), 
we may also freeze $t$ and we omit the variable $t$. We write 
$\|u\|_{\Hg}=\|u\|$. Since $A$ and $\ph$ are smooth and 
$\ph$ is bounded below it is well known 
(\cite{kato,Iwatsuka}) that 
$H_0$ on $C_0^\infty(X)$ is essentially selfadjoint; if we 
write the selfadjoint extension by the same letter, then 
$H_0$ is the maximal operator:
\[
D(H_0)=\{u \in \Hg \colon 
H_0 u \in \Hg\}
\]
where $H_0$ on the right hand side should be understood 
in the sense of distributions. It then is obvious that 
$\Si(2) \subset D(H_0)$. We prove the opposite inclusion and 
$H_0 \geq (Nd+1)/2$. Let $u \in D(H_0)$  
and take a sequence $u_n \in C_0^\infty(\R^n)$ such that 
$\|u_n -u\|\to 0$ and $\|H_0 u_n -H_0 u\|\to 0$ as 
$n\to \infty$. Then, 
$|\pa_j |u_n|| \leq |(\pa_j -iA_j)u_n|$ and 
property (a) above imply  
\[
(H_0 u_n, u_n) \geq \frac12 \|\nabla |u_n|\|^2 + 
\frac12\|\la\ux\ra^2  u_n \|^2 \geq 
\left(\frac{Nd+1}{2}\right) 
\|u_n\|^2. 
\]
Letting $n\to \infty$,  we obtain $H_0 \geq (Nd+1)/2$. 
Set $f=H_0 u$ and $f_n= H_0 u_n$. 
Then, Kato's inequality implies in the sense of distributions  
\[
\left(-\frac12 \lap + \frac12\la\ux\ra^2 \right)|u_n|
\geq |f_n|.
\]
Then, the property (c) above of the harmonic oscillator 
implies  
\bqn \lbeq(step-1)
|u_n(\ux)|\leq \left(H_{os} + \frac12 \right)^{-1}|f_n|
\leq 2\la \ux\ra^{-2}|f_n|(\ux).
\eqn 
Expanding $(-i\nabla_j-A(x_j))^2$ in $f_n = H_0 u_n $, 
we see that  
\bqn \lbeq(u-1)
\left(-\frac12\lap  + \frac12\ux^2\right)u_n  
= f_n + i\sum_{j=1}^N A(x_j)\cdot \nabla_j u_n
+ O(\ux^2)u_n. 
\eqn 
We want to show that 
\bqn \lbeq(key)
\|A(x_j)\cdot \nabla_j u_n\|\leq C \|f_n\|, 
\quad j=1,\dots, N, 
\eqn 
which, by virtue of \refeq(u-1) and \refeq(step-1), will imply 
\[
\|H_{os}u_n\|\leq  \|f_n\| + C\|f_n\| + \|O(\ux^2)u_n\|
\leq C \|f_n\|.  
\]
Then, the standard argument will imply $u \in D(H_{os})$ 
and $\|H_{os}u\|\leq C\|f\|$ and, complete the proof of statement (1). 
To show \refeq(key), we solve \refeq(u-1) for $u_n$ 
and apply $\pa_{jk}x_{jl}$, $k, l=1, \dots, d$ to 
the resulting equation:
\bqn \lbeq(step-0)
u_n = H_{os}^{-1}\left(f_n+ 
i\sum_{j=1}^N A(x_j)\cdot \nabla_j u_n+O(\ux^2)u_n\right). 
\eqn 
Then, it is clear from property (b) of $H_{os}$ 
and \refeq(step-1) that 
\[
\|\pa_{jk}x_{jl}H_{os}^{-1}(f_n+O(\ux^2)u_n)\|
\leq C \|f_n+O(\ux^2)u_n\|\leq C\|f_n\|.
\]
In the right of 
\begin{multline*}
\pa_{jk}x_{jm}H_{os}^{-1} \pa_{ln} A_n(x_l)u_n 
=\pa_{jk}H_{os}^{-1}[H_{os}, x_{jm}] 
H_{os}^{-1} \pa_{ln} A_n(x_l)u_n \\
+ \pa_{jk}H_{os}^{-1}[x_{jm}, \pa_{ln}] A_n(x_l)u_n
+ \pa_{jk}H_{os}^{-1} \pa_{ln}x_{jm}A_n(x_l)u_n
\end{multline*}
we have $[H_{os}, x_{jm}]= -\pa_{jm}$ and 
$[x_{jm},\pa_{ln}]= -\delta_{jm,ln}$. 
It follows by virtue of property (b) of $H_{os}^{-1}$ 
and \refeq(step-1) again that 
$\|\pa_{jk}x_{jm} H_{os}^{-1} \pa_{ln} A_n(x_l)u_n \|
\leq C \|f_n\|$, hence 
$\|x_{jm} \pa_{jk} H_{os}^{-1} \pa_{ln} A_n(x_l)u_n \|
\leq C \|f_n\|$. The desired estimate \refeq(key)  
follows evidently. This proves the statement (1). 

By the assumptions on $A$ and $\ph$, 
$t \to H_0(t)u\in \Hg$ for $u \in \Si(2)$ is differentiable and 
\[
\dot H_0(t) u = \left(i\sum_{j=1}^N \dot A(t,x_j)\nabla_j + 
i\frac12 {\rm div}_j\, \dot A(t,x_j) + A(t,x_j)\dot A(t,x_j)+ 
\dot \ph(t,x_j)\right)u  
\]
is continuous with values in $\Hg$. Statement (2) is now 
obvious . 
\edpf 

\subsection{Proof of \refthb(2)}

The proof is an improvement of that of Theorem 7 
of \cite{Y-2}. Define for compact intervals $I$ 
the pair of function spaces 
\begin{gather} 
\Yg(I)= \{u \in C(I, \Si(2)) \colon \dot u \in \Xg(I)\}, \\
\Yg^\ast (I)= \{u \in C(I, \Hg) \colon 
\dot u \in \Xg^\ast(I)\}. 
\end{gather}
They are Banach spaces with natural norms 
\begin{gather} 
\|u\|_{\Yg(I)}= \|u\|_{C(I, \Si(2))}+ \|\dot u \|_{\Xg(I)}, \\
\|u\|_{\Yg^\ast (I)}= \|u\|_{C(I, \Hg)}+  
\|\dot u\|_{\Xg^\ast(I)}.
\end{gather}

The following identities will be frequently used in what 
follows.  
\bglm For $f\in \Si(2)$ and $f \in \Hg$ respectively, 
we have identities  
\begin{align} \lbeq(id-1)
& H_0(t) U_0(t,s)f = U_0(t,s) H_0(s)f
+ \int_s^t U_0(t,r) \dot H_0(r) U_0(r,s)f dr. \\ 
& U_0(t,s)H_0(s)^{-1}f= H_0(t)^{-1}U_0(t,s)f  \notag \\
& \hspace{4cm} 
- \int_s^t U_0(t,r)
\left( \frac{d}{dr} H_0(r)^{-1}\right) U_0(r,s)f dr.
\lbeq(id-2)
\end{align}
\edlm 
\bgpf Let $f, g \in \Sg(X)$. Since $H_0(t)$ is selfadjoint 
in $\Hg$, we have 
\[
\frac{d}{dr}(H_0(r)U_0(r,s)f, U_0(r,t)g)= 
(\dot H_0(r)U_0(r,s)f, U_0(r,t)g).
\]
Integrate both sides by $r$ from $s$ to $t$ and obtain  
\[
(H_0(t)U_0(t,s)f, g)- (H_0(s)f,U_0(s,t)g) 
= \int^t_s (\dot H_0(r)U_0(r,s)f, U_0(r,t)g)dr, 
\]
which proves \refeq(id-1) on $\Sg(X)$. Since both 
sides of \refeq(id-1) is bounded from $\Si(2)$ to $\Hg$, 
it holds for $f \in \Si(2)$. Proof for 
\refeq(id-2) is similar and is omitted. 
\edpf 

\bglm \lblm(la-1) Let $f\in \Si(2)$ and $u_0(t)=U_0(t,s)f$.   
Then, for any compact interval $I\subset I_0$, $u_0 \in \Yg(I)$ and 
$\|u_0\|_{\Yg(I)}\leq C \|f\|_{\Si(2)}$ 
for a $C>0$ independent of $f$ and $I$.
\edlm 
\bgpf As was remarked under \refeq(d-propj) that 
$u_0 \in C(\R, \Si(2))$. By virtue of \refeq(id-1), we have 
\[
i\frac{d}{dt} u_0(t)= U_0(t,s) H_0(s) f + 
\int_s^t U_0(t,r) \dot H_0(r) U_0(r,s)f dr 
\equiv u_1(t)+ u_2(t). 
\]
Then, \refeq(stri-1) implies 
$\|u_1\|_{\Xg(I)}\leq C\|H_0(s)f\|_{\Hg}\leq C \|f\|_{\Si(2)}$. 
By virtue of \reflm(res-h) (2), 
$\dot H_0(r) U_0(r,s)f$ is a continuous function of 
$r$ with values in $\Hg$ and 
$\|\dot H_0(r) U_0(r,s)f\|_{L^1(I, \Hg)}\leq C \|f\|_{\Si(2)}$. 
It follows from \refeq(L1-stri) that 
\[
\|u_2\|_{\Xg(I)}\leq C \|f\|_{\Si(2)}.
\] 
This proves the lemma. 
\edpf 

Recall that $\Gg_s$ is the integral operator 
defined by \refeq(gsdef). 

\bglm \lblm(Ggs)
Let $s\in I_0$ and $I=[s-L, s+L]\subset I_0$. Then, 
$\Gg_s\in \Bb(\Yg^\ast(I), \Yg(I))$ and, 
for a constant $C>0$ independent of $s$ and $L$ 
\bqn \lbeq(yest)
\|\Gg_s u\|_{\Yg(I)} \leq C \|u\|_{\Yg^\ast(I)}, \quad u \in \Yg^\ast(I). 
\eqn 
\edlm 
\bgpf It suffices to show the following two estimates:
\bqn \lbeq(two)
\|\Gg_s u\|_{C(I, \Si(2))} \leq C \|u\|_{\Yg^\ast(I)}, \quad  
\|\dot \Gg_s u\|_{\Xg(I)} \leq C \|u\|_{\Yg^\ast(I)}.
\eqn 
Let $u \in \Yg^\ast (I)$. Then $u(r)$ is AC on $I$ 
with values in $\Si(-2)$ and $H_0(r)^{-1}$ is 
$C^1$ with values in $\Bb(\Si(-2), \Hg)$ by virtue 
of \reflm(res-h) (2). It follows that 
$U_0(t,r)H_0(r)^{-1}u(r)$ is AC with respect to 
$r \in I$ with values in $\Hg$. 
We differentiate this function and integrate 
by $r$ from $s$ to $t$. This gives the following 
integration by parts formula:   
\begin{align}
& \Gg_s u(t)=- \int_s^t (\pa_r U_0(t,r))H_0(r)^{-1}u(r) dr 
\notag \\
& = U_0(t,s)H_0(s)^{-1} u(s) - H_0(t)^{-1}u(t) \lbeq(bdry) \\ 
&+ \int_s^t U_0(t,r)(\pa_r H_0(r)^{-1}) u(r) dr 
+ \int_s^t U_0(t,r)H_0(r)^{-1}\dot u(r) dr. \lbeq(int-term)
\end{align}
Since $H_{0}(t)^{-1} \in \Bb(\Hg, \Si(2))$ is strongly 
$C^1$ function of $t$ by (2) of \reflm(res-h), 
it is obvious that two terms on \refeq(bdry) and 
the first integrals are $\Si(2)$-valued continuous 
functions of $t\in I$ and their norm in 
$C(I,\Si(2))$ are bounded by 
$C\|u\|_{C(I,\Hg)}\leq C\|u\|_{\Yg^\ast(I)}$. 
Define for $u \in \Yg^\ast (I)$ 
\bqn \lbeq(wt)
w(t)= \int_s^t U_0(t,r)\dot u(r) dr.
\eqn 
Then, \reflm(stri) implies  
\bqn \lbeq(w)
w(t) \in \Xg(I), \quad 
\|w\|_{\Xg(I)}
\leq C \|\dot u\|_{\Xg^\ast(I)}
\leq C \|u\|_{\Yg^\ast(I)}.
\eqn 
In particular $\|w\|_{C(I, \Hg)}\leq C \|u\|_{\Yg^\ast(I)}$. 
If we use 
\refeq(id-2) and change the order of integration, then 
\begin{align}
& \int_s^t U_0(t,r)H_0(r)^{-1}\dot u(r) dr 
= H_0(t)^{-1} \int_s^t U_0(t,r)\dot u(r) dr \lbeq(7-1) \\
& - \int_{s}^t U_0(t,\r)\pa_r (H_0(\r)^{-1})
\left(\int_{s}^\r U_0(\r,r)\dot u(r)  dr\right) d\r \\
& = H_0(t)^{-1}w(t)- 
\int_{s}^t U_0(t,\r)(\pa_r H_0(r)^{-1}) w(\r)d\r. 
\lbeq(7-2)
\end{align}
Since $w\in C(I, \Hg)$, \reflm(res-h) implies 
that two functions on \refeq(7-2) are both in  
$C(I,\Si(2))$ and are bounded by $C\|w\|_{C(I,\Hg)}$. 
Thus first of \refeq(two) is proved. 

We next prove the second of \refeq(two). 
Differentiating \refeq(bdry) and \refeq(int-term), we have 
\begin{multline}
i\frac{d}{dt}\Gg_s u(t) 
= H_0(t)U_0(t,s) H_0^{-1}(s) u(s) 
 + \int^t_s H_0(t) U_0(t,r)H_0(r)^{-1}\dot u(r) dr 
\\
- \int^t_s H_0(t) U_0(t,r)H_0(r)^{-1}\dot H_0(r) 
H_0(r)^{-1}u(r)dr\equiv a(t)+ b(t) - c(t), \lbeq(gsd-3)
\end{multline}
where the definition of $a(t), b(t)$ and $c(t)$ should be 
obvious. We rewrite these functions by using \refeq(id-1) 
and \refeq(id-2) in such a way that propagators $U_0(t,s)$ 
or $U(t,r)$ are placed on the left most or in front of 
$\dot u(r)$. Using \refeq(id-1) we rewrite: 
\begin{multline}
a(t)= U_0(t,s) u(s) \\ + 
\int_s^t U_0(t,r) \dot H_0(r) U_0(r,s)H_0(s)^{-1}u(s)dr 
=v_1(t) + v_2(t).
\end{multline}
\reflm(stri) implies 
$\|v_1\|_{\Xg(I)}\leq C \|u\|_{C(I, \Hg)} 
\leq C \|u\|_{\Yg^\ast (I)}$; 
\reflm(res-h) (2) implies 
$I \ni r \to
 \dot H_0(r) U_0(r,s)H_0(s)^{-1}u(s)\in \Hg$ 
is continuous and bounded by $C \|u\|_{\Yg^\ast (I)}$. 
It follows via \reflm(L1-stri) that  
$\|v_2\|_{\Xg(I)}\leq C \|u\|_{\Yg^\ast (I)}$. Hence 
\bqn \lbeq(a)
\|a\|_{\Xg(I)}\leq C \|u\|_{\Yg^\ast (I)}.
\eqn  
We rewrite $b(t)$ by using  
\refeq(id-1) to place $U_0(t,r)$ in the front, yielding  
\begin{multline*}
b(t)=\int^t_s U_0(t,r)\dot u(r) dr \\
+ \int^t_s \left(\int^t_r U_0(t,\r)\dot H_0(\r) U_0(\r,r)d\r\right) 
H_0(r)^{-1}\dot u(r) dr = v_3(t) + v_4(t).
\end{multline*}
We have $v_3(t)=w(t)$ and 
$\|v_3\|_{\Xg(I)}\leq C \|u\|_{\Yg^\ast(I)}$.  
Changing the order of integration yields  
\[
v_4(t)= \int_s^t U_0(t,\r) \dot H_0(\r)\left(
\int^\r_s U_0(\r, r) H_0(r)^{-1} \dot u(r)dr\right) d\r.
\]
Rewrite $U_0(\r, r) H_0(r)^{-1} \dot u(r)$ in 
the inner integral via \refeq(id-2) 
and change the order of integration in the 
resuting equation. We obtain  
\begin{align*} 
& v_4(t)= \int_s^t U_0(t,\r) \dot H_0(\r) H_0(\r)^{-1} \left(
\int^\r_s U_0(\r, r)\dot u(r)dr\right) d\r \\
& - \int_s^t U_0(t,\r) \dot H_0(\r) \left\{
\int^\r_s U_0(\r, \c)\pa_\c (H_0(\c)^{-1}) 
\left(\int_s^\c U_0(\c,r)\dot u(r)dr\right) d\c\right\} d\r 
\\ 
& \qquad = 
\int_s^t U_0(t,\r) \dot H_0(\r) H_0(\r)^{-1}w(\r)d\r \\
& \hspace{2cm} -\int_s^t U_0(t,\r) \dot H_0(\r) 
\left(\int^\r_s U_0(\r, \c)\pa_\c (H_0(\c)^{-1}) w(\c)
 d\c\right) d\r.
\end{align*}
By virtue of \refeq(w) and \reflm(res-h) (2), integrands 
on the right are both continuous functions 
of $\r$ with values in $\Hg$ and satisfy 
\begin{gather*}
\|\dot H_0(\r) H_0(\r)^{-1}w(\r)\|_{C(I, \Hg)}
\leq C\|u\|_{\Yg^\ast(I)}, \\
\left\|\dot H_0(\r) \left(\int^\r_s U_0(\r, \c)\pa_\c (H_0(\c)^{-1}) w(\c)d\c\right)
\right\|_{C(I, \Hg)}\leq C \|u\|_{\Yg^\ast(I)}.
\end{gather*}
Then, \refeq(L1-stri) once more produces
$\|v_4\|_{\Xg(I)}\leq C \|u\|_{\Yg^\ast(I)}$ and we obtain 
\bqn \lbeq(b) 
\|b\|_{\Xg(I)} \leq C \|u\|_{\Yg^\ast(I)}. 
\eqn 
Finally, we estimate $c(t)$. We first 
rewrite $H_0(t)U_0(t,r)$ by using \refeq(id-1). 
After changing the order of integration, we have 
\begin{align*}
& c(t)= \int^t_s U_0(t,r) \dot H_0(r) H_0(r)^{-1}u(r)dr  \\
& + \int^t_s U_0(t,\r)\left(
\int_s^\r  \dot H_0(\r) U_0(\r, r) H_0(r)^{-1}
\dot H_0(r) H_0(r)^{-1}u(r)dr
\right) d\r. 
\end{align*}
\reflm(res-h) (2) implies that integrands on 
the right are $\Hg$-valued continuous 
functions of $r$ and $\r$ respectively and 
\begin{gather*}
\|\dot H_0(r) H_0(r)^{-1}u(r)\|_{C(I, \Hg)}\leq 
C \|u\|_{\Yg^\ast(I)}, \\
\left\| \int_s^\r  \dot H_0(\r) U_0(\r, r) H_0(r)^{-1}
\dot H_0(r) H_0(r)^{-1}u(r)dr\right\|_{C(I, \Hg)}
\leq C \|u\|_{\Yg^\ast(I)}.
\end{gather*}
It follows from \reflm(L1-stri) that 
\bqn \lbeq(c)
\|c\|_{\Xg(I)}\leq C\|u\|_{\Yg^\ast(I)}. 
\eqn 
Combining \refeq(a), \refeq(b) and \refeq(c) with 
\refeq(gsd-3), we obtain the second estimate of 
\refeq(two), completing the proof of the lemma. 
\edpf 

\bglm \lblm(V-2)
Suppose $V$ satisfies \refass(V-2). 
Then, for any $\ep>0$ there exists $\d>0$  
such that for intervals of $|I|<\d$ there exists a constant 
$C_{\ep, I}$ such that 
\bqn 
\|\Vg u\|_{\Yg^\ast(I)}\leq \ep \|u\|_{\Yg(I)} 
+ C_{\ep, I} \|u\|_{C(I, \Hg)}, \quad u \in \Yg(I). 
\eqn 
\edlm  
\bgpf By virtue of \refeq(est-V) 
and the argument at the beginning of the proof of 
\refth(propa-1), there exists $\d>0$ such that we have  
for intervals of $|I|<\d$  that 
\bqn 
\|\Vg \dot{u} \|_{\Xg^{\ast}(I)}
\leq C \max_D \|V_D\|_{\Ig^{cont,p_D}_D(I)} 
\|\dot u \|_{\Xg(I)}
\leq \ep \|\dot u\|_{\Xg(I)}.
\lbeq(al)
\eqn 
If we write 
$\pa_t V_D= W_1+ W_2  \in 
L^{b_{D}}(I, L^{q_{D}}(X_{D,r})) + L^1(I, L^\infty(X_{D,r}))$ 
as in \refass(V-2), then 
Sobolev embedding theorem and H\"older's inequality 
imply that for sufficiently small $\d>0$ 
\begin{gather}
\|W_1 u \|_{L^{\th_D'}(I, L^{l_D',2}_D)}
\leq \|W\|_{L^{b_{D}}(I, L^{q_{D}}(X_{D,r}))} 
\|u\|_{C(I,\Si(2))} <\ep \|u\|_{\Yg(I)}. \lbeq(bet) \\
\|W_2 u\|_{L^1(I, L^2(X))}
\leq \|W_2 \|_{L^1(I,L^\infty(X_{D,r}))} 
\|u\|_{C(I,\Si(2))} <\ep \|u\|_{\Yg(I)}. \lbeq(gam)
\end{gather} 
Here we used that indices satisfy 
\[
\frac1{q_D}+ \left(\frac12-\frac2{n_D}\right) 
=\frac1{l_D'}, \quad 
\frac{1}{\th_D'}=\frac1{b_D}. 
\]
Combination of \refeq(al), \refeq(bet) and \refeq(gam) 
proves that, when $|I|<\delta$,  
\bqn \lbeq(-last)
\|(d/dt)(\Vg u)\|_{\Xg^\ast(I)}\leq \ep\|u\|_{\Yg(I)}.
\eqn 
When $V_D \in \Ig_D^{cont,\tilde p_D}(I)$, it is obvious that 
$V_D u \in C(I, \Hg)$ for $u\in C(I, \Si(2))$. We want show that 
\bqn \lbeq(last)
\|V_D(t,x_{D,r})u \|_{C(I,\Hg)}\leq 
\ep \|u\|_{C(I, \Si(2))} + C_\ep \|u\|_{C(I, \Hg)} 
\eqn 
for all $D \subset \{1,\dots, N\}$. The argument at the beginning of the 
proof of \refth(propa-1) once more shows that, 
for any $\ep>0$, we may write 
$V_D \in \Ig_D^{cont,\tilde p_D}(I)$ as 
\[
V_D =V_D^{(1)}+ V_D^{(2)}, \ \mbox{with}\ 
\|V_D^{(1)}\|_{C(I, L^{\tilde p_D}(X_{D,r}))}\leq\ep.
\]
Then, recalling that $\tilde{p}\geq 2$ if $n_D =3$, 
we obtain \refeq(last) by using H\"older's inequality and 
the Sobolev embedding theorem. Thus, $\|\Vg u\|_{C(I,\Hg)}\leq 
\ep \|u\|_{\Yg(I)} + C_\ep \|u\|_{C(I, \Hg)}$. 
This with \refeq(-last) proves the lemma. 
\edpf 

\paragraph{Completion of the proof of \refthb(2)} 
We let $f \in \Si(2)$ and $u(t)$ be the solution of the 
integral equation \refeq(int-eq):
\bqn \lbeq(int-q) 
u(t) = u_0(t) + (\Gg_s \Vg u)(t),\quad  u_0(t) = U_0(t,s) f.
\eqn 
It suffices to show that, when $L>0$ is sufficiently 
small, $u \in \Yg(I)$ for 
$I= [s-L, s+L]$. By virtue of \reflm(Ggs) and \reflm(V-2), 
we may take $L$ small so that 
\bqn \lbeq(ggsv)
\|\Gg_s \Vg\|_{\Xg(I)}<1/2, \quad 
\|\Gg_s \Vg u\|_{\Yg(I)}\leq (1/2)\|u\|_{\Yg(I)} 
+ C \|u\|_{\Xg(I)}.
\eqn
By virtue of \reflm(la-1) we have $u_0 \in \Yg(I)$ and, 
if we successively define
\[
u_n(t)=u_0(t) + \Gg_s \Vg u_{n-1},  \quad n=1,2, \dots,  
\]
then, \refeq(ggsv) implies  
$u_n \in \Xg(I)\cap \Yg(I)$ for $n=1,2, \dots$ and  
\begin{gather}
\|u_n-u_{n-1}\|_{\Xg(I)} \leq 2^{-n} \|u_0\|_{\Xg(I)}, \\
\|u_{n+1}-u_{n}\|_{\Yg(I)} \leq 
2^{-1}\|u_{n}-u_{n-1}\|_{\Yg(I)} 
+ C \|u_{n}-u_{n-1}\|_{\Xg(I)}.
\end{gather}
It follows that $u_n$ converges to the solution 
$u \in \Xg(I)$ and that  
\begin{align*}
\|u_{n+1}-u_{n}\|_{\Yg(I)}& \leq (1/2)\|u_{n}-u_{n-1}\|_{\Yg(I)} 
+ C 2^{-n}\|u_0\|_{\Xg(I)} \notag \\ 
& \leq 2^{-n}\|u_1-u_0\|_{\Yg(I)} 
+ C n 2^{-n} \|u_0\|_{\Xg(I)}. 
\end{align*}
Thus, $u_n$ converges to $u$ also in $\Yg(I)$. 
This proves $u \in C(I,\Si(2)) \cap C^1(I,\Hg)$.
The rest of the  theorem immediately follows 
from this and the proof of \refth(2) is completed. 
\qed

\end{document}